\documentclass[aps, showpacs, reprint,superscriptaddress]{revtex4-1}
\usepackage{graphicx,amsfonts,float}
\usepackage{epsf}
\usepackage{pstricks}
\usepackage{hyperref}

\begin{document}
  \title{Thermodynamics of non-Markovian reservoirs and heat engines}
 \author{George Thomas}
 \email{georget@imsc.res.in}
 \affiliation{Optics and Quantum Information Group, 
The Institute of Mathematical Sciences, HBNI,  CIT Campus, Taramani, Chennai 600113, India
}

\author{Nana Siddharth}
\email{nana.siddharth@gmail.com }
 \affiliation{ 
The Institute of Mathematical Sciences, HBNI,  CIT Campus, Taramani, Chennai 600113, India
}

\author{Subhashish Banerjee}
\email{subhashish@iitj.ac.in }
 \affiliation{
Indian Institute of Technology Jodhpur, Jodhpur 342011, India}

 \author{Sibasish Ghosh}
 \email{sibasish@imsc.res.in }
  \affiliation{Optics and Quantum Information Group, 
The Institute of Mathematical Sciences, HBNI,  CIT Campus, Taramani, Chennai 600113, India
}

\begin{abstract}
We show that  non-Markovian effects of the reservoirs can be used as a resource 
to extract work from an Otto cycle. 
The state transformation under
non-Markovian dynamics is achieved via a two-step process, namely an isothermal process using a Markovian reservoir
followed by an adiabatic process. From second law of thermodynamics, we show that the maximum amount of extractable work
from the  state prepared under the non-Markovian dynamics quantifies a lower bound of non-Markovianity. 
We  illustrate our ideas with an explicit example of non-Markovian evolution.
\end{abstract}

\maketitle
\section{Introduction}
The second law of thermodynamics is one of the unshaken pillars of 
modern physics.  The Kelvin-Planck statement of the second law of thermodynamics states that the
work cannot be extracted from a single bath (or different baths with same temperature) in a cyclic manner.  Maxwell introduced the idea of a demon which can create
a nonequilibrium state from an equilibrium state to extract work \cite{Leff-Book,Maruyama2009}.
The extraction of work from a single bath with the help of the demon can also be modeled using single molecule Szilard engine \cite{Szilard1929}.
This long-term puzzle of extracting work from a single bath in a cyclic manner, was solved by Landauer by
showing that true cyclicity is achieved only when the demon's memory is also erased. Landauer's erasure principle showed that the erasure of one bit
of memory increases the entropy of the universe by at least $k_B \ln {2}$, where $k_B$ is the Boltzmann constant \cite{Leff-Book,Landauer1961,Bennett1982,Kim2011}.
This remarkable statement brings out an intriguing connection between information and thermodynamics in
both classical and quantum regimes.

Another direction of attempts had been made to understand the validity of the second law
in the quantum regime. It has been shown that there can be an apparent violation of second law in different heat cycles
if we use quantum features such as  negentropy \cite{Scully2001}, 
 quantum correlations such as entanglement \cite{Scully2001,Allahverdyan2001,Allahverdyan2002,Lutz2009}, coherences \cite{Scully2003}, and squeezed thermal bath \cite{Lutz2014}.
 But once the preparation costs for these
quantum features are included, then these apparent violations vanish \cite{Zubairy2002, Hilt2011}. However, a
more general framework of second law of thermodynamics may be needed while  working in the quantum 
regime \cite{Brandao2015} or involving frictional effects \cite{Bizarro2012,Bizarro2015}.

At this point, it is interesting to study the extension of thermodynamics to quantum systems where the environmental 
effects are important \cite{Breuer-Book}. When the system is in contact with an environment, the dynamics of the system
may not be unitary and one needs the approach of open quantum systems. The interaction between the system and bath
can affect each other and in many cases, the Markov approximation is valid, which states that  the environment recovers  instantaneously as 
compared to the large timescale associated with the dynamics of the system, resulting in  continuous
flow of information from the system to the environment \cite{Alonso2017}. However, there are
many scenarios where the Markov approximation is not valid \cite{grabert88,sb1,sb2}
and a back flow of information from the environment to the system can take place.
These are the typical signatures of non-Markovian evolution \cite{Breuer-Book,Alonso2017,pradeep1,pradeep2}.

This work is focused on the properties of quantum heat engines mainly under non-Markovian baths.
Properties of heat engines under Markovian environment have been well studied in the past \cite{Alicki1979}.
The thermodynamic perspective of non-Markovian effects has attracted wide interest in recent years 
 \cite{Gelbwaser-Klimovsky2013a,Zhang2014a,Pekola2015,Bylicka2016,Omar2016,Chen2016,Whitney2016,Raja2017,Avijit2017,Rezakhani2017,Chen2017}.
Non-Markovian effects can be used to extract work from a single bath  \cite{Gelbwaser-Klimovsky2013a}
or can be used to exceed classical Carnot bound \cite{Zhang2014a}, showing an apparent violation of the second law. 
These apparent violations  are due to the memory present in the bath. There 
are also connections among emergence of non-Markovianity, system-environment interaction and violation of Landauer's bound \cite{Omar2016}.
 Moreover, even if a non-Markovian evolution  relaxes the system to  an equilibrium state,
the equilibrium state may not be invariant under time evolution \cite{Saverio2010,Rezakhani2017,Alonso2017}.
Negative entropy production rate due to non-Markovian dynamics drives the system  away
from  equilibrium \cite{Avijit2017}. The  system-environment correlations can take a part of the environment out of equilibrium \cite{Pekola2015}. 
The memory associated with a non-Markovian bath can have an
information theoretic interpretation and it can drive the system away from equilibrium.
But the second law of thermodynamics suggests that there is always a cost  to create a nonequilibrium state from an equilibrium state.

In this work, we use  non-Markovian effects which drive the system away from equilibrium, as a resource to
extract work from an Otto cycle. Understanding the thermodynamics of  non-Markovian evolution for which the equilibrium state is not an invariant state
is a nontrivial task since the integrated entropy production can also be negative \cite{Rezakhani2017}.
For such an evolution, we estimate the minimum thermodynamic cost for the non-Markovianity.
By accounting  the minimum cost for  non-Markovian effects, the Otto cycle becomes equivalent to a Carnot cycle.
Different models of quantum Otto cycle have been studied widely in recent years \cite{Kieu2004,Quan2007,Lutz2014, Zhang2014a, Thomas2011,Kosloff2015, Thomas2017,Asoka2017,Venu2017}.
Otto cycle is generally an irreversible cycle with efficiency less than the Carnot value. The main source of  irreversibility
is due to the temperature difference between the system and the bath during the thermalization processes. 
If the temperature gradient approaches zero in the starting of a thermalization process, then the efficiency approaches
Carnot value while the work output goes to zero. On the other hand, we show that using specific non-Markovian dynamics, 
one can extract nonzero work without irreversible entropy production in thermalization process. In this case, the maximum achievable
efficiency is Carnot efficiency.
Moreover, this work extraction is a finite-time process and hence it has a finite power output.
But the bath correlations may not vanish in a finite-time,  even though the system comes back to its initial state after each cycle.

The paper is organized as follows: In Section \ref{sec2}, 
we introduce the four-staged Otto cycle.
In Section \ref{sec3}, 
we discuss heat and work
in terms of relative entropies and further show that work can be extracted
even when the temperatures of the non-Markovian baths are equal.
We solve this apparent contradiction (with Kelvin-Planck statement of second law) by redefining the heat exchanged between the system and the baths.
We illustrate these ideas with an example of a non-Markovian dynamics in Section \ref{sec4}. 
Section \ref{sec5} is devoted to conclusions
and future directions.
\section{Model}
\label{sec2}
\begin{figure}[H]
\begin{center}
\includegraphics[width=8cm]{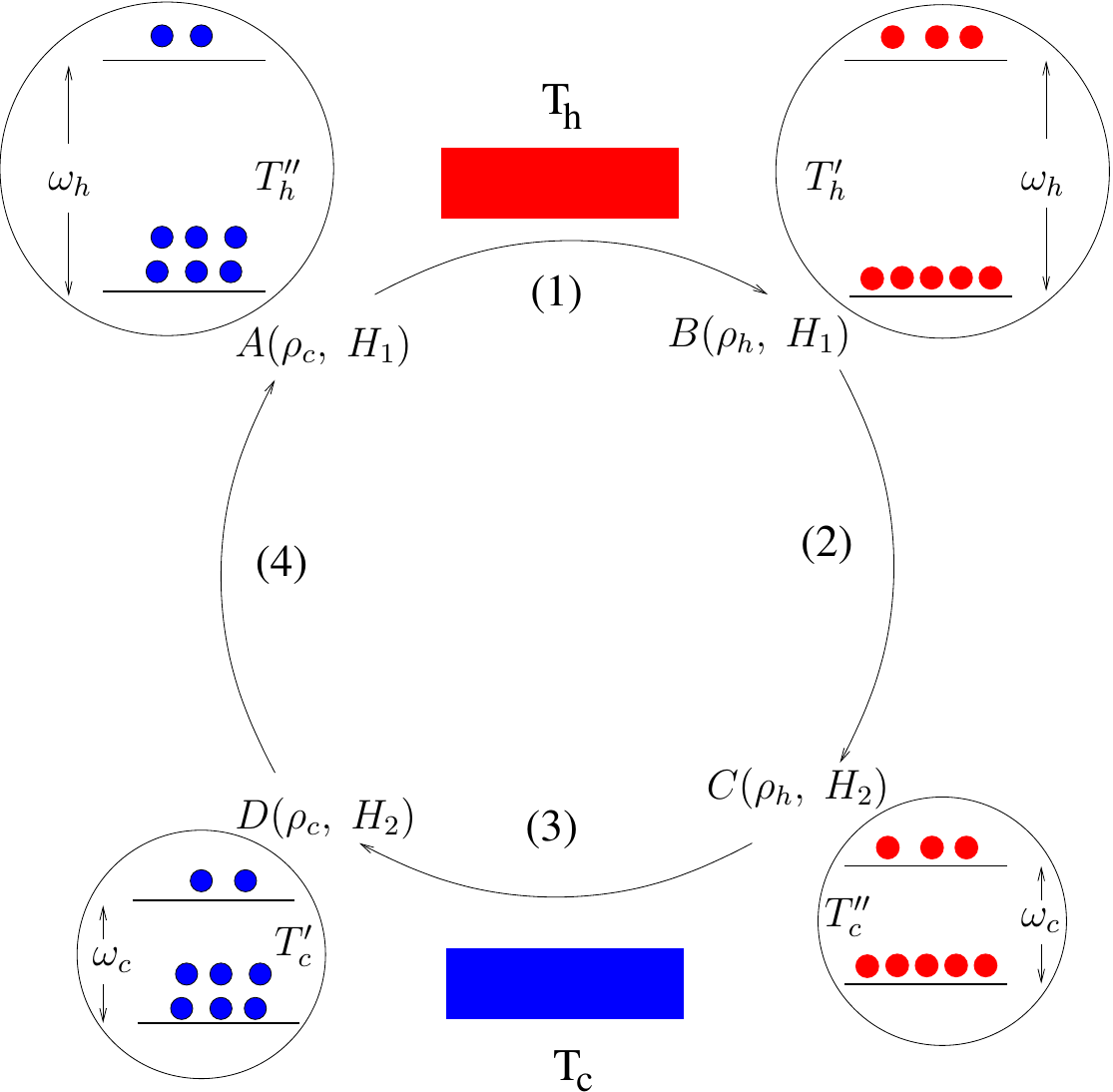}
\caption{A schematic representation of the quantum Otto cycle. In Stages 1 and 3, the system is in contact with the hot and the cold reservoir, respectively. 
The reservoirs can be Markovian or non-Markovian. See the main text for details about the symbols used.}
\hfill
\label{Otto_picture}
\end{center}
\end{figure}

The conventional classical Otto cycle consists of two isochoric (constant volume)  processes  and two
adiabatic processes and the working medium is taken to be an ideal gas.
Heat is exchanged with the baths during the isochoric processes and work is done during the adiabatic processes.
In the quantum model of an Otto cycle, the working medium is a quantum system, say a spin-1/2 system  \cite{Kieu2004,Quan2007}. 
Analogously to constant volume process in a classical Otto engine, for a two-level system, the energy-level
spacing is fixed during the quantum isochoric process.
During the isochoric process, the system is connected to a reservoir and hence we need  the approach
of open quantum system to study the evolution of the system. Depending on the nature of the reservoir,
the evolution can be Markovian or non-Markovian. In this section, we discuss a general framework of a four-staged Otto cycle
irrespective of Markovian or non-Markovian reservoirs. Further, in Section \ref{sec3},
we study the implications of using non-Markovian reservoirs
in a four-staged Otto cycle 
 and illustrate these ideas with an explicit example of non-Markovian evolution in Section \ref{sec4}.
Fig. \ref{Otto_picture} pictorially represents four stages of an Otto cycle, 
with a spin-1/2 system as the working medium.

Let us consider a two-level system  with energy eigenvalues $\omega/2$ and $-\omega/2$.
If the density matrix of the system is a thermal state corresponding to a temperature $T$,
then the canonical probability for the system being
in excited state is 
$P=1/[1+ \exp{(\omega/T)}]$ with ground state probability $1-P$, where we set Boltzmann constant $k_B=1$. 
The mean energy of the system is 
$(2P-1)\omega/2 = -(\omega/2) \tanh{(\omega/2T)}$.
In general, for  a two-level system, when the density matrix is diagonal in the eigenbasis of the Hamiltonian,
an effective temperature can be defined such that the probability distributions are canonical.
Let us briefly review a four staged quantum Otto cycle \cite{Kieu2004,Quan2007,Thomas2011,Thomas2017}.
\begin{itemize}
 \item 
\textit{Stage 1}: Consider a spin-$1/2$ system, prepared in  a density matrix $\rho_c$, diagonal
 in the eigenbasis of the Hamiltonian $H_1=(\omega_h/2)\sigma_z$, where $\omega_h=\gamma_0 B$ and $\sigma_z$ is the Pauli matrix. Here
$\gamma_0$ is a constant and $B$ is the constant magnetic field applied in the $z$-direction on the system. 
In this stage, the system is 
attached to a hot bath of temperature $T_h$ for a time $\tau_1$. 
The final state of the system is $\rho_h$, which is diagonal
in the eigenbasis of $H_1$. 
The effective temperature of the system at the end of the process is $T_h'=-\omega_h/(2\tanh^{-1}\langle \sigma_z \rangle_h)$, where
$\langle \sigma_z \rangle_h={\rm Tr}[\sigma_z \rho_h]$, i.e., $\rho_h=\exp{(-H_1/T_h')}/{\rm Tr}[{\exp{(-H_1/T_h')}}]$.
The heat absorbed from the hot bath  is $Q_h= {\rm Tr}[H_1(\rho_h-\rho_c)]$. 
 \item \textit{Stage 2}: The system is then decoupled from the bath. The magnetic field is varied from $\omega_h$ 
to $\omega_c$ adiabatically such that the final Hamiltonian at the end of Stage 2 is $H_2= (\omega_c/2)\sigma_z$,
while the initial Hamiltonian was $H_1= (\omega_h/2)\sigma_z$ ($\omega_c<\omega_h$). The density matrix ($\rho_h$) of the system 
remains unchanged throughout this process.
The work done in this process is equal to the change in mean energy $W_1= {\rm Tr}[\rho_h(H_1-H_2)]$. Due to work done by the system,
the temperature of the system changes;  at the end of this process it becomes  $T_c''=T_h'\omega_c/\omega_h$.
 \item \textit{Stage 3}: The system is then attached to the cold bath at temperature $T_c$, for a time $\tau_2$. The Hamiltonian is fixed
at $H_2$ during the process. After a time $\tau_2$, 
the system attains a state $\rho_c$ characterized by a temperature $T_c'=-\omega_c/(2\tanh^{-1}\langle \sigma_z \rangle_c)$, 
 where $\langle \sigma_z \rangle_c={\rm Tr}[\sigma_z \rho_c]$, in other words, 
 $\rho_c=\exp{(-H_2/T_c')}/{\rm Tr}[{\exp{(-H_2/T_c')}}]$. 
 The heat rejected to the cold bath in this case is $Q_c= {\rm Tr}[H_2(\rho_c-\rho_h)]$.
  \item  \textit{Stage 4}: The system is  detached from the cold bath and Hamiltonian is changed adiabatically from $H_2$ to $H_1$, keeping
  the density matrix ($\rho_c$) unchanged. 
 The work done in this process is $W_2= {\rm Tr}[\rho_c(H_2-H_1)]$. The temperature of the system at the end of the adiabatic process is given by
 $T_h''=T_c'\omega_h/\omega_c$.
 \end{itemize}
 In principle, the Stages 2 and 4 can be achieved instantaneously since $\langle \sigma_z \rangle_h$ and $\langle \sigma_z \rangle_c$ are constants of motion
 in the respective stages \cite{Geva1992}.
 In Stage 1 and Stage 3, the respective asymptotic state is an equilibrium state, i.e., $\lim_{\tau_1\to \infty}T_h'=T_h$
 and $\lim_{\tau_2\to \infty}T_c'=T_c$.  
The amount of heat that the system  absorbed from the  hot bath after a finite time $\tau_1$ is given by
\begin{eqnarray}
Q_h =\frac{\omega_h}{2} \left({\tanh}\left[\frac{\omega_c}{2T_c'}\right]-\tanh\left[\frac{\omega_h}{2T_h'}\right]\right).
\label{Qh_single_spin}
\end{eqnarray}
Similarly, the amount of heat  rejected to the cold bath in a finite time $\tau_2$ is given as
\begin{eqnarray}
Q_c =-\frac{\omega_c}{2} \left({\tanh}\left[\frac{\omega_c}{2 T_c'}\right]-\tanh\left[\frac{\omega_h}{2 T_h'}\right]\right).
\label{Qc_single_spin}
\end{eqnarray}
The work done by the system is $W=W_1+W_2=Q_h+Q_c$. Hence, we have
\begin{eqnarray}
W =\frac{(\omega_h-\omega_c)}{2} \left({\tanh}\left[\frac{\omega_c}{2 T_c'}\right]-\tanh\left[\frac{\omega_h}{2 T_h'}\right]\right).
\label{work_single_spin}
\end{eqnarray}
The eigenvalues of the Hamiltonian $H_1$ and $H_2$ are $(\omega_h/2,-\omega_h/2)$ and $(\omega_c/2,-\omega_c/2)$, respectively.
Therefore, the canonical probabilities for the system being in excited state at the end of Stage 1 and Stage 3 are $P_h=1/(1+\exp{[\omega_h/T_h']})$ 
and $P_c=1/(1+\exp{[\omega_c/T_c']})$, respectively. The  Eq. (\ref{work_single_spin}) can be rewritten in terms of the excited state probabilities as 
\begin{eqnarray}
W =(\omega_h-\omega_c) \left(\frac{1}{1+e^{\omega_h/ T_h'}}-\frac{1}{1+e^{\omega_c/ T_c'}}\right).
\label{work_single_spin_canonical}
\end{eqnarray}

For a machine to work as an engine, the working medium has to absorb heat from the hot reservoir, and a part of it has to be converted to useful work
by the system, and the remaining heat has to be rejected to the cold reservoir. Hence, we should have $W>0$, $Q_h>0$ and $Q_c<0$. This implies
\begin{equation}
\frac{\omega_c}{T_c'}\geq\frac{\omega_h}{T_h'}.
\label{condition}
\end{equation}
Therefore, the efficiency of the system is 
\begin{equation}
\eta=\frac{W}{Q_h}=1-\frac{\omega_c}{\omega_h}\leq1-\frac{T_c'}{T_h'}.
\end{equation}
When we use  Markovian dynamics in Stage 1 and Stage 3, to satisfy  Eq. (\ref{condition}), we should have $T_h>T_c$.  Let us try to explain it.
In order to absorb heat from the hot bath, the initial temperature of the system $T_h''$ in Stage 1 must satisfy $T_h''<T_h'$. Similarly, to reject heat to the cold bath
we should have $T_c''>T_c'$.
Depending upon Markovian and non-Markovian reservoirs, the effective  temperatures of the system
at a given time of the evolution may differ.
For a Markovian reservoir, the temperature of the system monotonically 
approaches the bath temperature, therefore we  have $T_h''<T_h'<T_h$ and $T_c''>T_c'>T_c$. 
In other words, with Markovian reservoirs, the 
 maximum and minimum effective temperature attainable for the system are $T_h$ and $T_c$, respectively.
 Hence, we have
$T_c\leq T_c'\leq T_h'\leq T_h$.
Therefore, we have $\eta \leq 1-T_c'/T_h'\leq 1-T_c/T_h$ (Carnot efficiency), establishing  the consistency with the second law of thermodynamics.
 On the other hand, with non-Markovian effects, the temperature of the system can attain higher values than $T_h$ during the hot bath contact and similarly,
 during the cold bath contact, the temperature of the system can fall below $T_c$.
 This is because the temperature of the system may not approach
the temperature of the bath monotonically, and  we can have a scenario $T_c'\leq T_c\leq T_h\leq T_h'$. 
Hence, the  ratio of the effective temperatures decreases, giving the possibility for the system to attain an efficiency 
more than Carnot value.
Next, we show that  Eq. (\ref{condition}) is satisfied using non-Markovian reservoirs even when $T_h=T_c$.
\section{Heat engine with non-Markovian reservoirs}
\label{sec3}
Let us consider an example where the system in state $\rho(0)$ at time $t=0$ is attached to a bath of  temperature $T$.
After a time $t$, let the  state of the system be $\rho(t)$. 
The Hamiltonian $H$ of the system remains unchanged during the evolution. We consider the  dynamics in such a way that 
the system  asymptotically (in time) reaches the equilibrium state $\rho^{\rm eq}$, corresponding to the bath temperature $T$:
$\rho^{\rm eq}=\exp{(-H/T)}/{\rm Tr}[\exp{(-H/T)}]$.
The heat flow between the system and the bath is equal to the change in  mean energy of the system. 
Therefore, the heat exchanged between the system and the bath after a finite-time $t$ is
\begin{eqnarray}
Q&=&{\rm Tr} [\rho(t) H]-{\rm Tr} [\rho(0) H]\nonumber\\
&=&-T\; {\rm Tr}[\rho(t)\ln{\rho^{\rm eq}}]+ T\;{\rm Tr}[\rho(0)\ln{\rho^{\rm eq}}]\nonumber\\
&=&T [S(\rho(t)||\rho^{\rm eq})-S(\rho(0)||\rho^{\rm eq})]
+ T( S_{v}[\rho(t)]- S_{v}[\rho(0)]),\nonumber
\label{heat_entropy}\\
\end{eqnarray}
where the relative entropy is $S(A||B)=-{\rm Tr}[A \ln{B}]-S_{v}[A]$ and the von Neumann entropy is $S_{v}[A]=-{\rm Tr}[A \ln{A}]$.
In general, in an isochoric process, the system may not be in equilibrium with the bath during the evolution, i.e., $\rho(0)$ and $\rho(t)$
 are nonequilibrium states
 in Eq. (\ref{heat_entropy}). If the final state is the equilibrium state ($\rho(t)= \rho^{\rm eq}$),
then the quantity $S(\rho(t)||\rho^{\rm eq})+  S_{v}[\rho(t)]$ becomes $ S_{v}[\rho^{\rm eq}]$.
Suppose the dynamics of the system  corresponds to a completely positive, trace preserving (CPTP) map $\Phi_t$, 
then for any two density matrices $\rho_1$ and $\rho_2$, we have \cite{Lindblad1975,nielsen-book}
\begin{equation}
 S(\Phi_t \rho_1||\Phi_t \rho_2)\leq S(\rho_1||\rho_2).
\end{equation}
For a Markovian dynamics $\Phi_t$, we have (as $\Phi_t \rho^{\rm eq} = \rho^{\rm eq}$, for all $t$)
\begin{equation}
S(\Phi_t\rho(0)||\Phi_t\rho^{\rm eq})=S(\rho(t)||\rho^{\rm eq})\leq S(\rho(0)||\rho^{\rm eq}),
\label{mark_cond_entropy}
\end{equation}
a result that also follows from Spohn's theorem \cite{Spohn1978}.
So we have $S(\rho(t)||\rho^{\rm eq})- S(\rho(0)||\rho^{\rm eq})\equiv\Delta D<0$ for any Markovian dynamics. 
But for a non-Markovian dynamics, for which $\Phi_t\rho^{\rm eq} \ne \rho^{\rm eq}$, we can have $\Delta D>0$.
Now, let us consider the cycle described in Section \ref{sec2}. 
The heat exchanged between the system and the hot  and  cold baths (Stages 1 and 3), respectively, are 
\begin{eqnarray}
Q_h&=&T_h [S(\rho_h||\rho^{\rm eq}_h)-S(\rho_c||\rho^{\rm eq}_h)]+ T_h \Delta S_v,
\label{Qh}\\
Q_c&=&T_c [S(\rho_c||\rho^{\rm eq}_c)-S(\rho_h||\rho^{\rm eq}_c)]-T_c \Delta S_v,
\label{Qc}
\end{eqnarray}
where $\Delta S_{v}=S_v(\rho_h)-S_v(\rho_c)$. Therefore, the work done is
\begin{eqnarray}
W&=&T_h [S(\rho_h||\rho^{\rm eq}_h)-S(\rho_c||\rho^{\rm eq}_h)]\nonumber\\
&+& T_c [S(\rho_c||\rho^{\rm eq}_c)-S(\rho_h||\rho^{\rm eq}_c)] \nonumber\\
&+&(T_h-T_c) [S_v(\rho_h)-S_v(\rho_c)].
\label{work}
\end{eqnarray}
When $T_h=T_c=T$, we have
\begin{eqnarray}
W=T [S(\rho_h||\rho^{\rm eq}_h)-S(\rho_c||\rho^{\rm eq}_h)+S(\rho_c||\rho^{\rm eq}_c)-S(\rho_h||\rho^{\rm eq}_c)].\nonumber\\
\label{work_same_T}
\end{eqnarray}
For Markovian dynamics, from Eq. (\ref{mark_cond_entropy}),
we have $W<0$, i.e., no work can be extracted when the temperatures of the baths are equal. This is because $\Phi_t\rho^{\rm eq}_h = \rho^{\rm eq}_h$
and $\Phi_t'\rho^{\rm eq}_c = \rho^{\rm eq}_c$, where $\Phi_t$ and $\Phi_t'$ are the maps corresponds to the evolution of the system
during Stage 1 and Stage 3, respectively. Therefore, from the properties of CPTP map, we  
have $S(\rho_h||\rho^{\rm eq}_h)-S(\rho_c||\rho^{\rm eq}_h)<0$. Similarly, we also have $S(\rho_c||\rho^{\rm eq}_c)-S(\rho_h||\rho^{\rm eq}_c)<0$
and hence the total work given in Eq. (\ref{work_same_T}) becomes negative, or the sole effect of the cycle is that the net work  is done on the system. On the other hand,
 considering a class of non-Markovian dynamics, for which the equilibrium state is not an invariant state, we have $\Phi_t\rho^{\rm eq}_h \ne \rho^{\rm eq}_h$
and $\Phi_t'\rho^{\rm eq}_c \ne \rho^{\rm eq}_c$. Hence we can have 
$S(\rho_h||\rho^{\rm eq}_h)-S(\rho_c||\rho^{\rm eq}_h)>0$ and $S(\rho_c||\rho^{\rm eq}_c)-S(\rho_h||\rho^{\rm eq}_c)>0$.
Therefore, for non-Markovian dynamics, we may get a positive work, implying an apparent violation
of the second law. Below, we will argue that there is no violation of the second law of thermodynamics
in our set up, once we define the actual amount of heat exchanged by non-Markovian heat baths by considering
the cost for non-Markovianity.

To characterize non-Markovian effects, we further simplify this problem, by attaching the system, which is in a thermal state 
at temperature 
$T_h$ ( $T_c$), to the hot (cold) bath at temperature $T_h$ ( $T_c$) at the beginning of the thermalization process. 
In this case, the states do not evolve under  Markovian dynamics as they are already in equilibrium with the baths.
On the other hand, the system may evolve when coupled to a non-Markovian 
bath.
Therefore, the heat exchanged between the system and the bath is solely due to the non-Markovian dynamics.
In our model of the heat cycle, we can prepare the system in the
equilibrium state, before attaching with the baths if we have  $\rho_c=\rho^{\rm eq}_h$ (for
attaching with the hot bath) and $\rho_h=\rho^{\rm eq}_c$ (for attaching with the cold bath). Then we can rewrite Eqs. (\ref{Qh}-\ref{work}) as,
\begin{eqnarray}
Q_h&=&T_h S(\rho^{\rm eq}_c||\rho^{\rm eq}_h)+ T_h [S_v(\rho^{\rm eq}_c)-S_v(\rho^{\rm eq}_h)],
\label{Qh1}\\
Q_c&=&T_c S(\rho^{\rm eq}_h||\rho^{\rm eq}_c)- T_c [S_v(\rho^{\rm eq}_c)-S_v(\rho^{\rm eq}_h)],
\label{Qc1}\\
W&=&Q_h+Q_c,
\label{work1}
\end{eqnarray}
since $S(\rho^{\rm eq}_c||\rho^{\rm eq}_c)=S(\rho^{\rm eq}_h||\rho^{\rm eq}_h)=0$.
At the end of Stage 1 (after a finite time $\tau_1$), the canonical probability of the system being in an
excited state is $P_h=1/(1+\exp{[\omega_h/T_h']})$ while at equilibrium  $P^{\rm eq}_h=1/(1+\exp{[\omega_h/T_h]})$. 
Similarly, at the end of Stage 3, the probability of the system to occupy the excited state is
$P_c=1/(1+\exp{[\omega_c/T_c']})$ while the equilibrium value is $P_c^{\rm eq}=1/(1+\exp{[\omega_c/T_c]})$.
The density matrix $\rho_h$ and $\rho^{\rm eq}_h$ are diagonal and the diagonal elements are ($P_h$, $1-P_h$) 
and ($P^{\rm eq}_h$, $1-P^{\rm eq}_h$), respectively. Similarly, the diagonal elements for $\rho_c$ and $\rho^{\rm eq}_c$
are ($P_c$, $1-P_c$) and ($P^{\rm eq}_c$, $1-P^{\rm eq}_c$), respectively.
Now $\rho_c=\rho^{\rm eq}_h$ and $\rho_h=\rho^{\rm eq}_c$, implies $P_c=P^{\rm eq}_h$ and $P_h=P^{\rm eq}_c$ and therefore we have
\begin{equation}
 \frac{\omega_c}{T_c'}=\frac{\omega_h}{T_h};\;\;\;\; \frac{\omega_h}{T_h'}=\frac{\omega_c}{T_c}.\label{noneqlconds}
\end{equation}
Eq. (\ref{noneqlconds}) satisfies the condition for engine, 
given in Eq. (\ref{condition}) when  $\omega_h/T_h\geq\omega_c/T_c$. In that case, the efficiency of the engine is
\begin{equation}
 \eta=1-\frac{\omega_c}{\omega_h}=1-\sqrt{\frac{T_c T_c'}{T_h T_h'}}.
\end{equation}
Under non-Markovian dynamics, the effective temperature of the system can oscillate around its equilibrium value and hence
it is possible to choose the values  of effective temperatures $T_h'$ and $T_c'$ such that $T_c'/T_h'<T_c/T_h$. In that case, the efficiency ($\eta$) can exceed the Carnot bound ($1-T_c/T_h$). 
This apparent violation is due to non-Markovian effects which can be ascribed to the memory of the bath. To analyze it,
let us assume that the system is decoupled  from the reservoir after a finite-time. We also assume that the decoupling process is sudden \cite{notes} and the  decoupling energy is negligible,
since the system is weakly coupled to the bath. At any instant of time, the state of the total system 
can be represented as the product of time-evolved state of the system and  the state of the bath (which is assumed to be unchanged),
upto the  second order in the strength of the interaction Hamiltonian \cite{Goan2011}. 
The total entropy change after decoupling the system from the hot bath  is
\begin{equation}
\Delta S_{\rm tot} = -\frac{Q_h}{T_h}+ [S_v(\rho^{\rm eq}_c)-S_v(\rho^{\rm eq}_h)].
\end{equation}
The first and the second term on the right hand side corresponds to entropy change 
in the reservoir and the system, respectively. Here, $Q_h=T_h[S(\rho^{\rm eq}_c||\rho^{\rm eq}_h)
+ S_v(\rho^{\rm eq}_c)-S_v(\rho^{\rm eq}_h)]$. One can see that $\Delta S_{\rm tot}$
for the non-Markovian dynamics is less than zero if we do not subtract at
least $T_h S(\rho^{\rm eq}_c||\rho^{\rm eq}_h)$ amount of energy from $Q_h$. Subtracting
$T_h S(\rho^{\rm eq}_c||\rho^{\rm eq}_h)$, from $Q_h$, we see that the total entropy production
at any instant of time is zero, similarly to the case of Markovian dynamics for the equilibrium initial state.
Therefore, the actual quantity for heat has to be defined as $\widetilde {Q_h}= Q_h-T_h S(\rho^{\rm eq}_c||\rho^{\rm eq}_h)$.

\begin{figure}[H]
\begin{center}
\includegraphics[width=6cm]{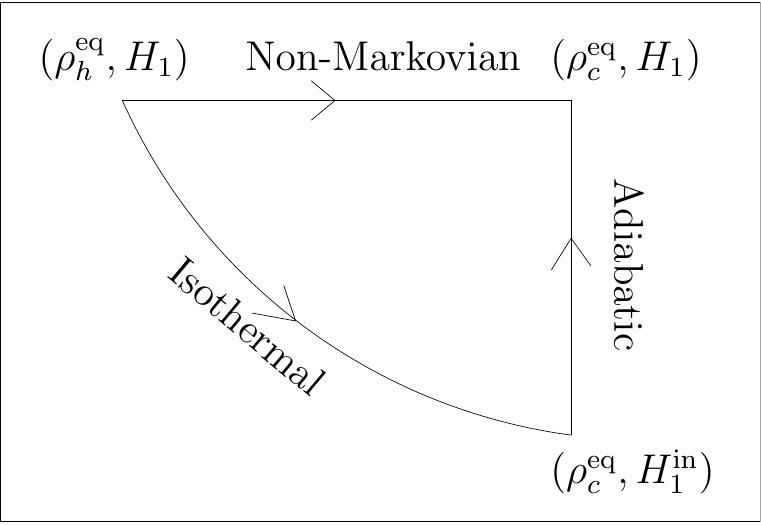}
\caption{A schematic representation of the state transformation under non-Markovian reservoir is achieved by a two-step protocol involving an isothermal process
using a Markovian reservoir followed by an adiabatic process. Here, $H_1^{\rm in}$ is the intermediate Hamiltonian.}
\hfill
\label{fig2step}
\end{center}
\end{figure}

For further understanding of the definition of  actual heat,
let us consider again the first isochoric process (Stage 1), where the Hamiltonian is fixed, but 
the state of the system is changed from $\rho^{\rm eq}_h$ to $\rho^{\rm eq}_c$ (see the discussion above Eq. (\ref{noneqlconds})).
The initial-to-final-state transition can be achieved as follows
with a two-step protocol involving a Markovian reservoir, as given in Fig. \ref{fig2step}.
\begin{itemize}
 \item To transform the state of the system from $\rho^{\rm eq}_h=\exp{(-H_1/T_h)}/{\rm Tr}[\exp{(-H_1/T_h)}]$ 
 to $\rho^{\rm eq}_c=\exp{(-H_1^{\rm in}/T_h)}/{\rm Tr}[\exp{(-H_1^{\rm in}/T_h)}]$, an isothermal process is carried out 
 by changing the Hamiltonian from $H_1$ to $H_1^{\rm in}$ and attaching the system with a Markovian bath of temperature $T_h$.
 The change in the mean energy is $\Delta U=\Delta F+ T_h [S_v(\rho^{\rm eq}_c)-S_v(\rho^{\rm eq}_h)]$.
 Here the work done is equal to the difference in free energies of two equilibrium states 
 $\Delta F=T_h \ln{({\rm Tr}[\exp{(-H_1/T_h)}]/{\rm Tr}[\exp{(-H_1^{\rm in}/T_h)}])}$ and the amount of heat exchanged is  $T_h[S_v(\rho^{\rm eq}_c)-S_v(\rho^{\rm eq}_h)]$.
 \item The system is then decoupled from the heat bath and thereby undergoes an adiabatic process where the Hamiltonian of the system 
 is changed to the initial Hamiltonian by keeping the density matrix fixed at $\rho^{\rm eq}_c$.
 The work done in this process is equal to the change in mean energy as there is no action of heat bath, and therefore, the work is given by,  ${\rm Tr}[\rho^{\rm eq}_c(H_1-H_1^{\rm in})]=T_h S(\rho^{\rm eq}_c||\rho^{\rm eq}_h)-\Delta F$.
\end{itemize}
Therefore, the total work done on the system to transform its state from 
$\rho^{\rm eq}_h$ to $\rho^{\rm eq}_c$  is  $T_h S(\rho^{\rm eq}_c||\rho^{\rm eq}_h)$
and the heat contribution is $T_h[S_v(\rho^{\rm eq}_c)-S_v(\rho^{\rm eq}_h)]$ \cite{Esposito2011}.
 The heat exchanged between the system and the bath 
is obtained by subtracting the amount of work done from the total change in the internal energy.
Now, coming back to the non-Markovian case, by subtracting the work-like term ($T_h S(\rho^{\rm eq}_c||\rho^{\rm eq}_h)$)
from $Q_h$ (given in Eq. (\ref{Qh1})), we get the actual heat exchanged as
$\widetilde {Q_h}= Q_h-T_h S(\rho^{\rm eq}_c||\rho^{\rm eq}_h)$.
With a similar argument, the actual heat exchanged with the cold reservoir (given in Eq. (\ref{Qc1})) is
$\widetilde {Q_c}= Q_c-T_c S(\rho^{\rm eq}_h||\rho^{\rm eq}_c)$.
Therefore, the work done by the engine after subtracting the minimum cost for non-Markovianity is
\begin{equation}
 \widetilde {W}=\widetilde {Q_h}+\widetilde {Q_c}=(T_h-T_c)[S_v(\rho^{\rm eq}_c)-S_v(\rho^{\rm eq}_h)],
 \label{work_final}
\end{equation}
and the efficiency is $\eta=\widetilde {W}/\widetilde {Q_h}=1-T_c/T_h$. As can be seen, when $T_h=T_c$, the work goes to zero,
which is consistent with the second law of thermodynamics. By accounting for the minimum costs of creating
the nonequilibrium states from the thermal equilibrium states, the Otto cycle constructed with non-Markovian reservoirs
becomes equivalent to a Carnot cycle with Markovian reservoirs, as shown in Fig. (\ref{fig1}). The description of the Carnot cycle in this context is
provided in the Appendix \ref{appendix_Carnot}. The temperature-entropy diagram for the cycle in shown in Fig. (\ref{fig1_TS}), where the equivalence of the Carnot and Otto cycles,
after considering the minimum cost for non-Markovianity, is depicted.
\begin{figure}[h]
\begin{center}
\includegraphics[width=6cm]{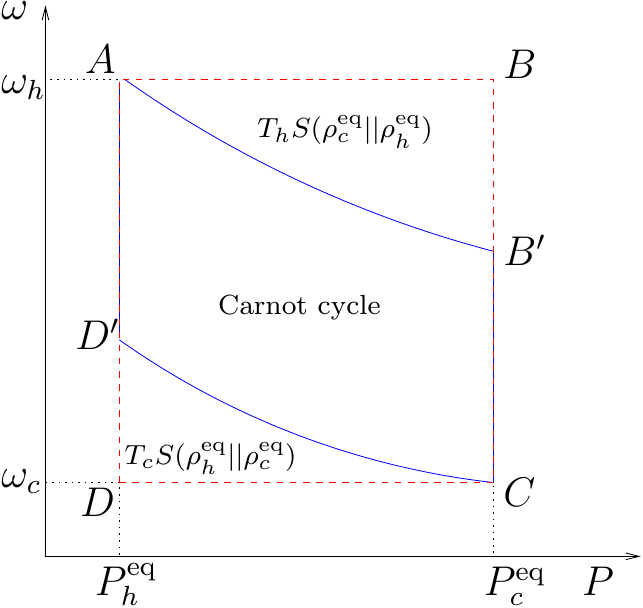}
\caption{Carnot cycle is embedded inside an Otto cycle. 
The cycle $ABCDA$ represents the Otto cycle while the cycle $AB'CD'A$ shows the Carnot cycle.
$\omega$ and $P$ are the energy level spacing and excited state population of the two-level system, respectively.
The efficiency of the Otto cycle can be greater than the Carnot efficiency.
The minimum work cost to create $\rho^{\rm eq}_c$ from $\rho^{\rm eq}_h$  
is $ T_h S(\rho^{\rm eq}_c||\rho^{\rm eq}_h)$ and the work cost to create 
$\rho^{\rm eq}_h$ from $\rho^{\rm eq}_c$  is $ T_c S(\rho^{\rm eq}_h||\rho^{\rm eq}_c)$. 
By accounting for these costs, the Otto cycle becomes equivalent to a Carnot 
cycle and the engine efficiency falls down to Carnot value. $P_c^{\rm eq}$ and $P_h^{\rm eq}$ are the excited state
populations at the end of Stage 1 and Stage 3 of the Otto cycle, respectively.
}
\hfill
\label{fig1}
\end{center}
\end{figure}

\begin{figure}[h]
\begin{center}
\includegraphics[width=6cm]{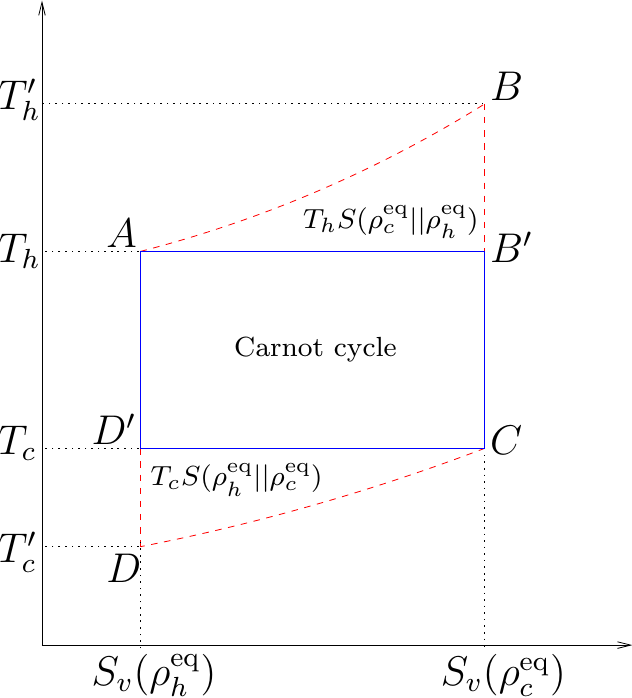}
\caption{Temperature-entropy diagram. Carnot cycle is embedded inside an Otto cycle. 
The cycle $ABCDA$ represents the Otto cycle while the cycle $AB'CD'A$ shows the Carnot cycle. The state of the system at point $B$ is a nonequilibrium
state for the bath at temperature $T_h$. Similarly, the state of the system at $D$ is nonequilibrium state for the bath at temperature $T_c$.
By accounting for the minimum costs for creating these nonequilibrium states, the Otto cycle becomes equivalent to a Carnot 
cycle.
}
\hfill
\label{fig1_TS}
\end{center}
\end{figure}
At any finite temperature difference of the baths,
the efficiency of the engine is the Carnot value. This can be  explained in terms of entropy production.
Before attaching the system to the cold or hot baths, its temperature is same as the bath temperature.
Thus, there is no irreversible entropy production due to the temperature 
gradient, unlike the usual Otto cycle. Therefore, one expects to obtain Carnot efficiency for this engine.

Another question at this point is the validity of the idea of quantifying non-Markovianity 
in terms of minimum cost of work  $T_hS(\rho^{\rm eq}_c||\rho^{\rm eq}_h)$ or $T_cS(\rho^{\rm eq}_h||\rho^{\rm eq}_c)$.
To understand this, let us consider an evolution of the system which is weakly coupled to a non-Markovian bath of temperature  $T$,
such that the system can evolve from an equilibrium state $\rho^{\rm eq}$
to $\rho'$ with a constant system Hamiltonian $H$. Subsequently, after decoupling from the non-Markovian bath, 
work can be extracted by transforming $\rho'$ back to $\rho^{\rm eq}$ using a Markovian bath and external driving. 
The maximum extractable work in this case is 
$TS(\rho'||\rho^{\rm eq})$ \cite{Esposito2011,Aberg2013,Popescu2014}. 
Therefore, the minimum energy cost to create the nonequilibrium state $\rho'$ using a non-Markovian dynamics is $TS(\rho'||\rho^{\rm eq})$.
Now, coming back to the case of engine, when the actual cost is more than $ T_h S(\rho^{\rm eq}_c||\rho^{\rm eq}_h)$ in Stage 1 or $ T_c S(\rho^{\rm eq}_h||\rho^{\rm eq}_c)$
in Stage 3, the efficiency of the engine constructed with non-Markovian reservoirs goes below the Carnot value. 
The aforesaid discussion holds good for any non-Markovian reservoir for which the asymptotic state of the system is the thermal
equilibrium state associated with the bath temperature, but not an invariant state.
\section{Non-Markovian dynamics}
\label{sec4}
In this section, we study a specific example of 
non-Markovian dynamics of a two-level quantum system \cite{Goan2011} to bolster the discussion in Section \ref{sec3}.
Let us consider a two-level system with energy
level spacing $\omega_A$ interacting with a reservoir of harmonic oscillators of frequencies $\{\omega_i\}$. The Hamiltonian
of the system plus bath can be written as
\begin{equation}
H=\frac{\omega_A}{2}\sigma_z+\sum_ig_i(\sigma_-a^{\dagger}_i+\sigma_+a_i)+\sum_i\omega_ia^{\dagger}_ia_i,
\end{equation}
where $a^{\dagger}_i$ and $a_i$ are the raising and lowering operators for the $i$th oscillator
and $g_i$ represents the interaction strength of the two level system with the $i$th oscillator. 
Here, $\sigma_+$ and $\sigma_-$ are the raising and lowering operators for the spin-1/2 system, respectively.
The evolution of  $\langle\sigma_z\rangle$, under weak coupling non-Markovian dynamics, is \cite{Goan2011}
\begin{eqnarray}
 \frac{d\left<\sigma_z(t)\right>}{dt} = &-&\left[ \Gamma_1(t) + \Gamma^*_1(t) + \Gamma_2^*(t) + \Gamma_2(t)\right]\left<\sigma_z(t)\right>\nonumber \\
 &-& \left[ \Gamma_1(t) + \Gamma^*_1(t) - \Gamma_2^*(t) - \Gamma_2(t)\right],
 \label{sigmaz}
\end{eqnarray}
while the off-diagonal terms evolves as
\begin{equation}
 \frac{d\left<\sigma_-(t)\right>}{dt} = -i\omega_A\left<\sigma_-(t)\right> -\left[ \Gamma_1(t) +
 \Gamma^*_2(t) \right]\left<\sigma_-(t)\right>\label{sigma-},
   \end{equation}
 \begin{equation}
   \frac{d\left<\sigma_+(t)\right>}{dt} = i\omega_A\left<\sigma_+(t)\right>
   +\left[ \Gamma^*_1(t) + \Gamma_2(t) \right]\left<\sigma_+(t)\right>\label{sigma+},
 \end{equation}
where we have
\begin{eqnarray}
 \Gamma_1(t) &=& {\displaystyle \int_0^t d\tau  \, \alpha_1(t-\tau) e^{i \omega_A(t-\tau)}},\\
 \Gamma_2(t)&=& {\displaystyle \int_0^t  d\tau \,  \alpha_2(t-\tau) e^{-i \omega_A(t-\tau)} }.
\end{eqnarray}
The environment correlations  functions $\alpha_1$ and $\alpha_2$  are defined as
\begin{eqnarray}
\alpha_1(t-\tau) &=& {\displaystyle \int_0^{\infty}d\omega \, \left(n(\omega) +1\right) J(\omega) e^{-i \omega(t-\tau) } }, \\
\alpha_2(t-\tau) &=& {\displaystyle \int_0^{\infty} d\omega \, n(\omega) J(\omega) e^{i \omega(t-\tau) } },
\end{eqnarray}
where $J(\omega)=\sum_i|g_i|^2\delta(\omega-\omega_i)=\gamma\omega \exp{(-\omega^2/\lambda^2)}$ 
is the spectral density and $\lambda$ is the cut-off frequency. Also, $n(\omega)=[\exp{(\beta \omega)}-1]^{-1}$ is the average number 
of  photons emitted with frequency $\omega$ from the reservoir at inverse temperature $\beta$. Here, $\gamma$  is a constant, characterizing the interaction
strength to the environment. 
In this dynamics, if we start with an initial state of the system which is diagonal in the energy eigenbasis,
 the off-diagonal elements of the system remain zero during the evolution.
We have $d\langle\sigma^\pm(t)\rangle/dt = P(t) \langle\sigma^\pm (t)\rangle $, 
where $ P(t)$ does not depend on $\langle\sigma_z(t)\rangle$, cf. Eqs. (\ref{sigma-}) and (\ref{sigma+}).
Thus, we can always associate an effective temperature 
with the two-level system.
\subsubsection*{Quantum heat engine}
\label{sec4.1}
\begin{figure}[h!]
\begin{center}
\includegraphics[width=9cm]{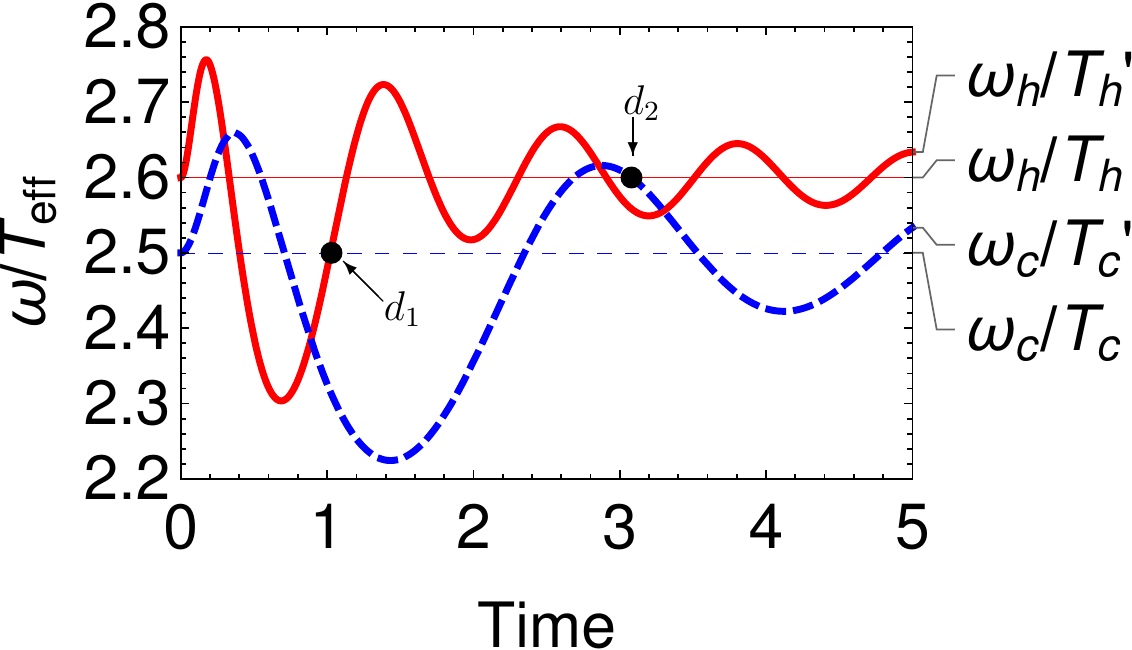}
\caption{(color online) The plot shows behavior of the ratio of energy level spacing ($\omega$) and effective temperature ($T_{\rm eff}$) versus time. 
The effective temperatures ($T_h'$ and $T_c'$) of the system is a function of time.
If we choose the time allocated for the thermalization branches
such that $\omega_h/ T_h'=\omega_c/ T_c$ and $\omega_c/ T_c'=\omega_h/ T_h$, then the
system works as an engine solely due to non-Markovian effects. One such pair of points are marked with dots ($d_1$ and $d_2$).
Here we use $\omega_c=2.5$, $\omega_h=5.2$, $\lambda=4 \omega_k (k=h,c)$, and $T_c=1$, $T_h=2$, and $\gamma=0.1$.
At a given time, $\omega_k/ T_k'= -2\tanh^{-1}[\langle \sigma_z \rangle_k]$,
where $\langle \sigma_z \rangle_k$ is obtained from Eq. (\ref{sigmaz}) of the corresponding bath. Such oscillations can be observed for a wide range of
parameters, $\gamma$ and $\lambda$.}
\label{fig2}
\end{center}
\end{figure}

With the above mentioned example of non-Markovian reservoirs, we can construct
a quantum heat cycle. 
In  the thermalization branches, 
we attach the system to the heat baths at temperature $T_h (T_c)$.
Because of the non-Markovian dynamics, the effective temperature oscillates around $T_h (T_c)$. For $\omega_h>\omega_c$, 
 we choose the time intervals 
 with which the system is attached to each bath   such that the condition for engine
(Eq. (\ref{condition})) is satisfied.
 Fig. \ref{fig2} depicts the curves corresponding to $\omega_h/ T_h'$ and $\omega_c/ T_c'$. 
 The thick solid  curve represents the evolution of $\omega_h/T_h'$ when the system is in contact with the hot bath (Stage 1). The initial value 
of this quantity is 
the equilibrium value (i.e., $\omega_h/T_h$) shown by the thin solid horizontal line. Due to non-Markovian dynamics, the effective temperature ($T_h'$) oscillates. 
When the value $\omega_h/T_h'=\omega_c/T_c$ (shown by the dot $d_1$), then the system is decoupled from the hot bath. In a similar way, 
the thick dashed  curve depicts the evolution of $\omega_c/T_c'$ when the system is coupled to the cold bath (Stage 3).
The thin dashed line represents the equilibrium value $\omega_c/T_c$.
When  $\omega_c/T_c'=\omega_h/T_h$ (denoted by the dot $d_2$), the system is decoupled from the cold bath.
 The resources for
 the work extraction are solely due to non-Markovian effects when the time allocated for the thermalization branches
 are such that $\omega_h/ T_h'=\omega_c/ T_c$ and $\omega_c/ T_c'=\omega_h/ T_h$ (Eqs. (\ref{condition}) and (\ref{noneqlconds})).  It is interesting to note that, 
 since the  initial temperatures of the system and bath are the same,
 there is no dissipation due to thermal gradients in the beginning of the thermalization process. 
 The efficiency of the engine in this case
 is $1-T_c/T_h$. The amount of extractable work, after deducting the minimum cost for non-Markovianity,
 is given in Eq. (\ref{work_final}). This work depends  on $\rho^{\rm eq}_h$ and $\rho^{\rm eq}_c$. 
 For  given $\omega_h$, $\omega_c$, $T_h$, and $T_c$, changing $\gamma$ or $\lambda$ can vary the characteristics of the oscillations (such as amplitude and frequency in Fig. \ref{fig2}), 
 but the amount of work given in Eq. (\ref{work_final}) does not change. But different values of $\gamma$ and $\lambda$  can change the time needed to achieve the 
 condition $\omega_h/ T_h'=\omega_c/ T_c$ and $\omega_c/ T_c'=\omega_h/ T_h$.
 Hence, the total time of cycle can also vary. It should be noted that
 the present model operates in the weak-coupling limit and thereby value of $\gamma$ is small.
\section{Conclusion and future directions}
\label{sec5}
We considered a four-staged Otto cycle with non-Markovian reservoirs.
The state of the system is the thermal state at the beginning of the isochoric process.
Hence, the heat exchanged during the process is solely due to the non-Markovian dynamics.
The system can extract work even when the temperatures of the baths are
equal, showing an apparent violation of second law of thermodynamics.
To resolve this apparent violation, the total energy difference in the system  due to non-Markovian bath is
decomposed into work-like and heat-like terms. Further, redefining the work and heat,
the efficiency of the engine takes the Carnot value and thus establishes
 consistency with the second law. We also show that the 
four-staged Otto cycle can be mapped to a four-staged Carnot cycle.
As a future direction, this work can be extended to coupled spins and coupled oscillators as working media for which
nonclassical features such as entanglement are also present \cite{Thomas2017}.
Our work can have a number of practical implications in modeling
engines working at finite power. 
The possibility of providing  operational quantifiers of non-Markovianity in terms of 
thermodynamic quantities can open up new avenues for the thermodynamics of open quantum systems.

Acknowledgement: G.T. thanks Ramandeep S. Johal for useful discussions.

\appendix 
 \section{Carnot cycle}
 \label{appendix_Carnot}
 Quantum Carnot cycle consists of two isothermal processes and two adiabatic processes, as shown in Fig. \ref{figure6n1} \cite{Quan2007}.
 The four stages are as follows:
 \begin{figure}[H]
\begin{center}
\includegraphics[width=5cm]{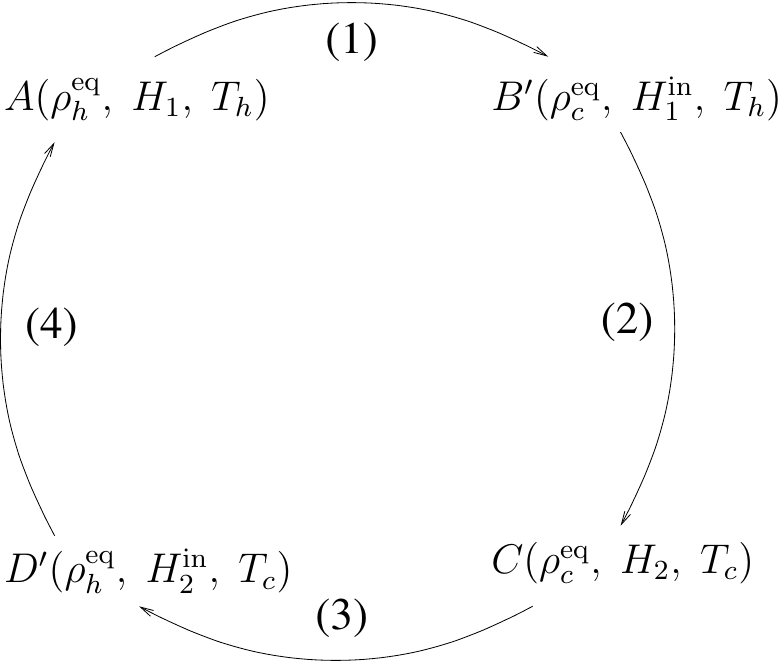}
\caption{A schematic representation of the quantum Carnot cycle.}
\label{figure6n1}
\end{center}
\end{figure}
 Stage 1: In this isothermal process, the Hamiltonian of the system is changed from $H_1=(\omega_h/2)\sigma_z$ to
 $H_1^{\rm in}=(\omega_h'/2)\sigma_z$ by changing the energy level spacing of the two-level system from $\omega_h$ to $\omega_h'$ while the system is in contact with the hot reservoir. 
 The temperature ($T_h$) of the system remains unchanged throughout Stage 1. The initial and the final density matrices of the system are given as
  $\rho^{\rm eq}_h=\exp{(-H_1/T_h)}/{\rm Tr}[{\exp{(-H_1/T_h)}}]$ and $\rho^{\rm eq}_c=\exp{(-H_1^{\rm in}/T_h)}/{\rm Tr}[{\exp{(-H_1^{\rm in}/T_h)}}]$, respectively.
 The heat absorbed from the hot bath contact is 
 \begin{eqnarray}
  Q_h^C&=&T_h [S_{\nu}(\rho^{\rm eq}_c)-S_{\nu}(\rho^{\rm eq}_h)]=\widetilde {Q_h}.
 \end{eqnarray}
 In terms of the energy level spacing and occupational probabilities, we can express the heat as
 \begin{eqnarray}
  Q_h^C&=&\frac{\omega_h}{\ln{k_h}} \left(P_c^{\rm eq}\ln{k_c}-P_h^{\rm eq}\ln{k_h}+\ln{\frac{1-P_h^{\rm eq}}{1-P_c^{\rm eq}}}\right),\nonumber\\
 \end{eqnarray}
 where $k_h=\ln{[(1-P_h^{\rm eq})/P_h^{\rm eq}]}$ and $k_c=\ln{[(1-P_c^{\rm eq})/P_c^{\rm eq}]}$. Here $P_h^{\rm eq}$ and $P_c^{\rm eq}$ are 
  probabilities of the excited state corresponds to the density matrices $\rho^{\rm eq}_h$ and $\rho^{\rm eq}_c$, respectively.
 Stage 2: The system is decoupled from the hot bath and the Hamiltonian is changed from   $H_1^{\rm in}$ to $H_2=(\omega_c/2) \sigma_c$ adiabatically.
  The density matrix ($\rho^{\rm eq}_c$) remains unchanged  throughout the process. In Stage 1, the energy level spacing $\omega_h'$ is chosen such that $\omega_h'/T_h=\omega_c/T_c$.
  Therefore, $\rho^{\rm eq}_c$ can  also be written as $\rho^{\rm eq}_c=\exp{(-H_2/T_c)}/{\rm Tr}[{\exp{(-H_2/T_c)}}]$. Thus the temperature of the system at the end of Stage 2 is $T_c$.
  Stage 3: The system is then attached to the cold bath. The Hamiltonian is changed isothermally from $H_2$ to $H_2^{\rm in}=(\omega_c'/2)\sigma_z$ ($\omega_c'=\omega_h T_c/T_h$).
  The initial and final
density matrices in this stage, are given as $\rho^{\rm eq}_c$ and $\rho^{\rm eq}_h$, respectively.
 Hence, the  heat rejected to the cold bath during  this isothermal process is 
  \begin{eqnarray}
  Q_c^C&=&-T_c [S_{\nu}(\rho^{\rm eq}_c)-S_{\nu}(\rho^{\rm eq}_h)]=\widetilde {Q_c}\\
  &=&-\frac{\omega_c}{\ln{k_c}} \left(P_c^{\rm eq}\ln{k_c}-P_h^{\rm eq}\ln{k_h}+\ln{\frac{1-P_h^{\rm eq}}{1-P_c^{\rm eq}}}\right).\nonumber\\
 \end{eqnarray}
 Stage 4: The system is decoupled from the system and the Hamiltonian of the system is adiabatically changed from $H_2^{\rm in}$ to $H_1$
 without altering the density matrix ($\rho^{\rm eq}_h$).
 The total work done by the system in the cycle is $W^C =Q_h^C+Q_c^C=\widetilde {W}$ and the efficiency is $\eta_c=1-T_c/T_h$.


\begin{thebibliography}{53}
\bibitem [{not()}]{notes}%
  \BibitemOpen
  \href@noop {} {}\bibinfo {note} {We are interested in the sudden process of
  decoupling because it does not change the state of the system. Therefore, we
  know the exact state of the system after decoupling. Moreover sudden
  decoupling reduces the total time of the cycle.}\BibitemShut {Stop}%
\end{thebibliography}
\end{document}